\documentclass{article}
\usepackage{spconf,amsmath,graphicx}
\usepackage[utf8]{inputenc}
\usepackage{cite}
\usepackage{physics}
\usepackage{placeins}
\usepackage{amssymb}
\usepackage{hyperref}
\usepackage{xcolor}
\usepackage{soul}
\usepackage{float}

\title{Quantum inspired image augmentation applicable to waveguides and optical image transfer via Anderson Localization}
\name{\begin{tabular}{c}Nikolaos E. Palaiodimopoulos$^{1}$, Vitor Fortes Rey$^{2}$, Matthias Tschöpe$^{2}$, Christina Jörg$^{3}$,\\
\textit{Paul Lukowicz$^{2,4}$, Maximilian Kiefer-Emmanouilidis$^{2, 3, 4}$}\end{tabular}}
\address{$^{1}$ Institute of Electronic Structure and Laser, FORTH, GR-70013 Heraklion, Greece \\
      $^{2}$ Embedded Intelligence, DFKI, Kaiserslautern D-67663, Germany\\
      $^{3}$ Department of Physics, RPTU Kaiserslautern-Landau, Kaiserslautern D-67663, Germany\\
      $^{4}$ Department of Computerscience, RPTU Kaiserslautern-Landau, Kaiserslautern D-67663, Germany}
\begin{document}
\ninept
\maketitle
\begin{abstract}
We present a quantum inspired image augmentation protocol which is applicable to classical images and, in principle, due to its known quantum formulation applicable to quantum systems and quantum machine learning in the future. The augmentation technique relies on the phenomenon Anderson localization. As we will illustrate by numerical examples the technique changes classical wave properties by interference effects resulting from scatterings at impurities in the material. We explain that the augmentation can be understood as multiplicative noise, which counter-intuitively averages out, by sampling over disorder realizations. Furthermore, we show how the augmentation can be implemented in arrays of disordered waveguides with direct implications for an efficient optical image transfer. 
\end{abstract}
\section{Introduction}
\label{sec:intro}
\begin{figure}[t]
     \centering
     \includegraphics[width=\columnwidth]{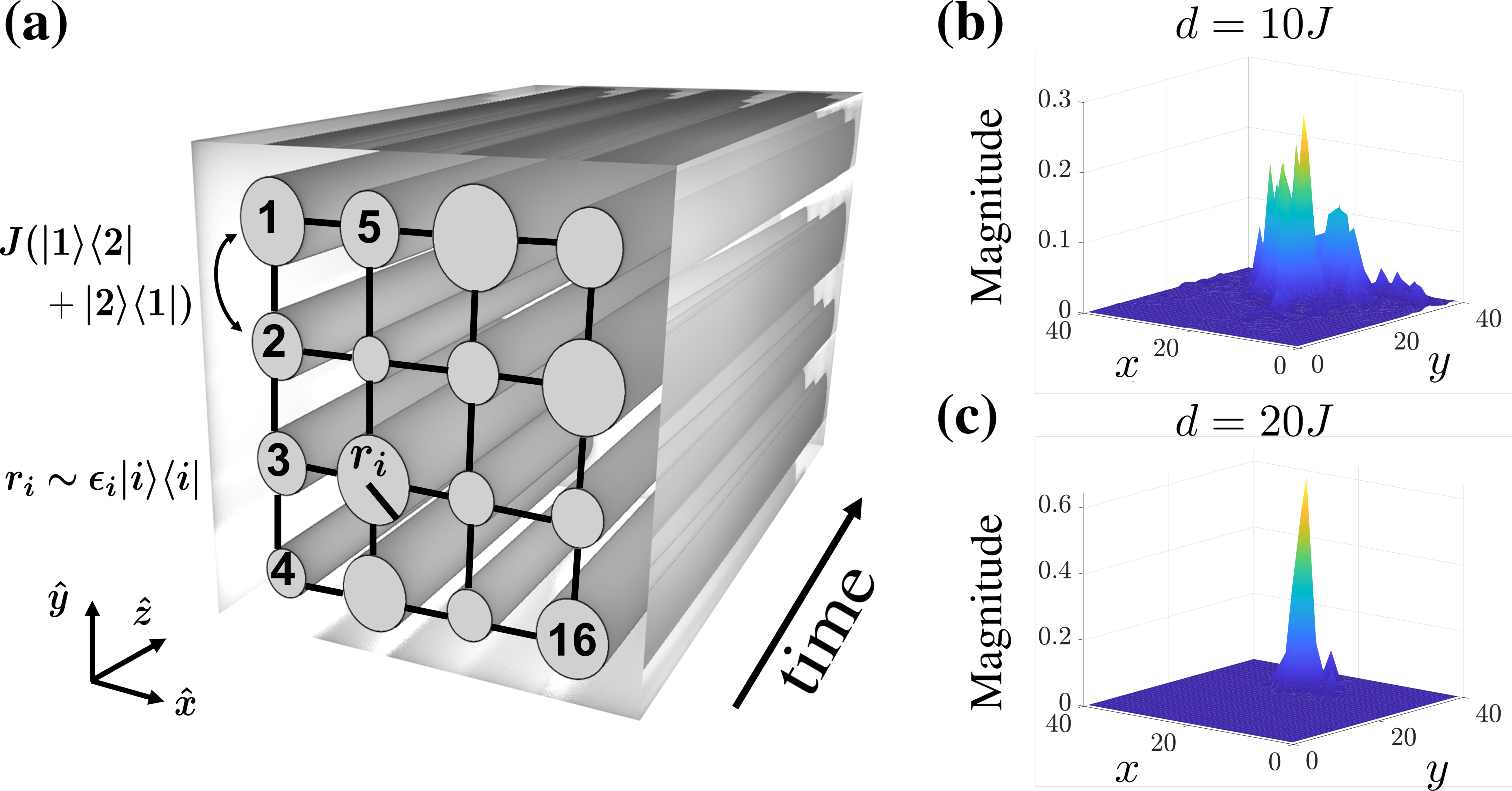}
     \caption{\textbf{(a)} Schematic of the waveguide setup arranged in a square-lattice geometry. The random radii are analogous to the onsite energies $\epsilon_{i}$ and we also depict an indicative example of one hopping process among the first two waveguides. \textbf{(b)},\textbf{(c)}: Two examples for the magnitudes of localized states extracted at $E_k=0$. For a disordered system $d=10J$ \textbf{(b)} one finds states with larger fluctuations of the magnitudes around the peak. \textbf{(c)} This behavior changes at much stronger disorders  $d=20J$. \vspace{-5pt} }
     \label{fig:lattice}
 \end{figure}

 Data augmentation \cite{van2001art} is a standard technique in machine learning, which increases the robustness of systems against input variations and facilitates  higher accuracy with a smaller number of labeled training data.  The fundamental reasons for the benefits of data augmentation are well understood and it is obvious that they remain valid when moving from classical to quantum machine learning\cite{Henderson2019, Mari2020, Chalumuri2022}. What is much less understood is how to best integrate data augmentation in quantum machine learning systems.  We first address this question by investigating how image augmentation can be performed as a quantum computing operation and applied to quantum encoded data, rather then a pre-processing step executed in a classical way applied to classically encoded data.  
 To understand the effects of the augmentation we will study an application of the quantum inspired protocols to the transfer of images through networks of waveguides. Sending an image as a collection of pixels via waveguides would be no problem if all waveguides are decoupled. However, if we want to use compact cables and hardware, arrays of waveguides have to be densely placed and they will be coupled evanescently, which leads to deterioration of the image after transfer. So how to decouple the waveguides? We focus on Anderson localization (AL)\cite{Anderson1958,AndersonLocalization}, a physical effect which, as explained in the following, is suited for classical and quantum image processing. AL is observed in waves which become localized in sufficiently disordered media, for example ultrasound in a disordered slab of aluminium spheres\cite{Hu2008, Goicoechea2020}, or light in optical waveguides with transverse spatially random refractive indexes or random radii\cite{Schwartz2007, Segev2013,Karbasi2014, Mafi2021}, see Fig.\ref{fig:lattice} (a). 
 AL emerges due to the interference of waves which are scattered from impurities. Counter-intuitively, the interference is only constructive at the initial point in space of the wave's formation and destructive everywhere else. This leads to a characteristic shape of the wave's magnitude decaying exponentially with distance to its initial point of formation, where the profile will not change in time. In practice, inelastic processes \cite{Scheffold2013} or unavoidable general dissipation dynamics will lead to decoherence in most uncontrolled cases \cite{Genway2014, Yusipov2017,Lorenzo2018}. This suppresses interference effects and finally destroys AL. Still, the  localized light in disordered optical fibers survive long enough, thus, AL is of high relevance for optical signal and image transfer in very dense structures\cite{Mafi2021}, effectively decoupling the waveguides. AL was initially understood in quantum systems, where non-interacting electrons become localized because of impurities within a metal\cite{Anderson1958, AndersonLocalization}. The localized behavior here emerges due to particle-wave duality and quantum interference of particles. With the rise of accessible, although noisy quantum hardware it was shown that signatures of AL can be observed in experiments \cite{Smith2019}. The possible applications in the future are vast, ranging from biomedical imaging\cite{Zhao2021} to quantum communication and imaging\cite{Demuth2022}. 

 In this paper we study the effects of a quantum inspired image protocol which uses Anderson localization and index permutations to augment initial images and achieves high quality image transfer. As we will show, the protocols have direct effect on the wave propagation and are suitable to augment data which inherits wave properties classically or quantum mechanically. It should be clearly stated that the augmentation protocol does not benefit from a quantum speed-up or advantage as long as classical images are used. This changes if the considered data or image inherits quantum properties\cite{Gresch2022} like entanglement, which allows for example for quantum enhanced sensing \cite{Conlon2023} or a quantum speed-up. While we do not consider entanglement based quantum enhancement here we show that our protocol is directly applicable to a system of optical waveguides, which allows a computation with speed of light of considered material and makes our approach also suitable for embedded systems and ubiquitous computing.

\section{Method}\label{sec:method}
Light propagation in waveguides is described by the paraxial Helmholtz equation, which is mathematically equivalent to the Schrödinger equation, describing the time evolution of wave functions in quantum-mechanical systems \cite{Szameit_2010, Giordani2021}. The propagation direction $z$ along the waveguide axis plays the role of time in the Schrödinger equation, while the refractive index landscape in waveguide arrays is replaced by the atomic potentials in condensed matter lattices. Thus, on-site potentials in the Hamiltonian can be tuned by either changing the waveguide radius or refractive index, and hopping with a constant $J$ corresponds to coupling of light between waveguides.
 Since we provide a protocol applicable to quantum systems, we describe the imaging protocol in the universal language of the Schrödinger's equation. For the reader not familiar with the Dirac bra-ket notation of quantum mechanics, we have prepared a brief introduction in the following. Up to a few general expressions, we will formulate our paper so that it can be understood in terms of simple linear algebra. 
To this end, let us first introduce briefly the Dirac bra-ket notation.
We note the following linear algebra analogies will hold for this paper. A ket-vector $\ket{j}$ is analog to unit vector $\hat{e}_{j}$ in an $N$-dimensional space, a hot encoded 1 at j-th index. A bra-vector $\bra{j}$ is then given by the transpose $\hat{e}_{j}^{T}$. Any general quantum state can be expressed by $\ket{\psi}=\sum_j c_j \ket{j}=\sum_j c_j \hat{e}_{j}$, where $c_j$ are the amplitudes and $|c_j|$ are the magnitudes of a state and $c_j\in \mathbb{C}$. The equivalent bra-vector is the complex conjugated ($*$ symbol) and transposed ($T$ symbol) ket-vector, where for both operations $*T\equiv\dagger$ is commonly used $\bra{\psi}=(\ket{\psi})^\dagger=\sum_j c^*_j \bra{j}=\sum_j c^*_j \hat{e}^T_{j}$. A scalar product of two vectors is given by $\bra{\phi}\ket{\psi}$, a norm by $N=\norm{\psi}=\sqrt{\bra{\psi}\ket{\psi}}$ and an outer product is given by $\ket{\phi}\bra{\psi}$. Furthermore, we note that in this paper inside bra or ket greek letters will indicate a general quantum state or special state and latin letters will correspond to unit vectors. Finally, a $m\times n$ matrix can be represented by $\mathbf{M}=\sum_{i=1,j=1}^{m,n} c_{ij} \ket{i}\bra{j}$. 

\subsection{Mapping an image to a quantum state}\label{Imageencoding}
Since we follow a quantum protocol we have to make sure that the image is also initially quantum encoded. Let us start from a grayscale picture, a $m \times n$ matrix $\mathbf{I}$, with elements $c_{ij} \in \left[0,255\right]$ corresponding to the grayscale value of each pixel
$\mathbf{I}= \sum_{i=1,j=1}^{m,n} c_{ij} \ket{i}\bra{j},$ with $c_{i,j}\in \mathbb{R}^+$.
To transform $\mathbf{I}$ into a proper quantum state we will use an amplitude encoding \cite{mottonen2004transformation, Weigold2022}. This means that we first have to flatten the matrix into a vector form or in other words stack the columns of a matrix $\mathbf{I}$ on top of each other creating a column vector with $m\cdot n$ components (this corresponds to a reshape in Fortran ordering $l=i+m(j-1)$). Furthermore, the state is normalized such that all vector components squared sum to 1, i.e. $\frac{1}{N}\sum_{l=1}^{m\cdot n}\abs{c_{l}}^2=1$, where $N=\sqrt{\sum_{l=1}^{m\cdot n}\abs{c_{l}}^2}$ is the norm. The general expression for an amplitude encoded image is then given by
\begin{align}
    \ket{\Psi_I}= \frac{1}{N} \sum_{l=1}^{m\cdot n} c_{l} \ket{l}.
    \label{Init State}
\end{align}
This will be the initial state for the protocol, where all amplitudes are real valued and positive and, thus, likewise magnitudes. In waveguide setups one would technically consider intensities, which are the squared magnitudes. Our protocol, though, works for both cases, whether we use magnitudes or intensities. Since all quantum operations are unitary and reversible the norm of the state stays intact. At any time the state can be rewritten as an image by reversing the previous mentioned steps, however, one has to keep in mind that the time-evolution will lead to complex valued coefficients (amplitudes). Thus, in a few words the procedure is as follows. Reshape the vector back into a matrix form $m\times n$ (Fortran ordering) and redo the normalization by multiplying the matrix by $N$. If the image was time-evolved the coefficients will be complex thus one has to take the magnitudes $\left|\bra{j} \ket{\psi}\right|$ to receive a proper picture. For simplicity, in the following we will only consider square images of size $n\times n$. 

{\ninept
\begin{figure*}[!ht]
\includegraphics[width=1\textwidth]{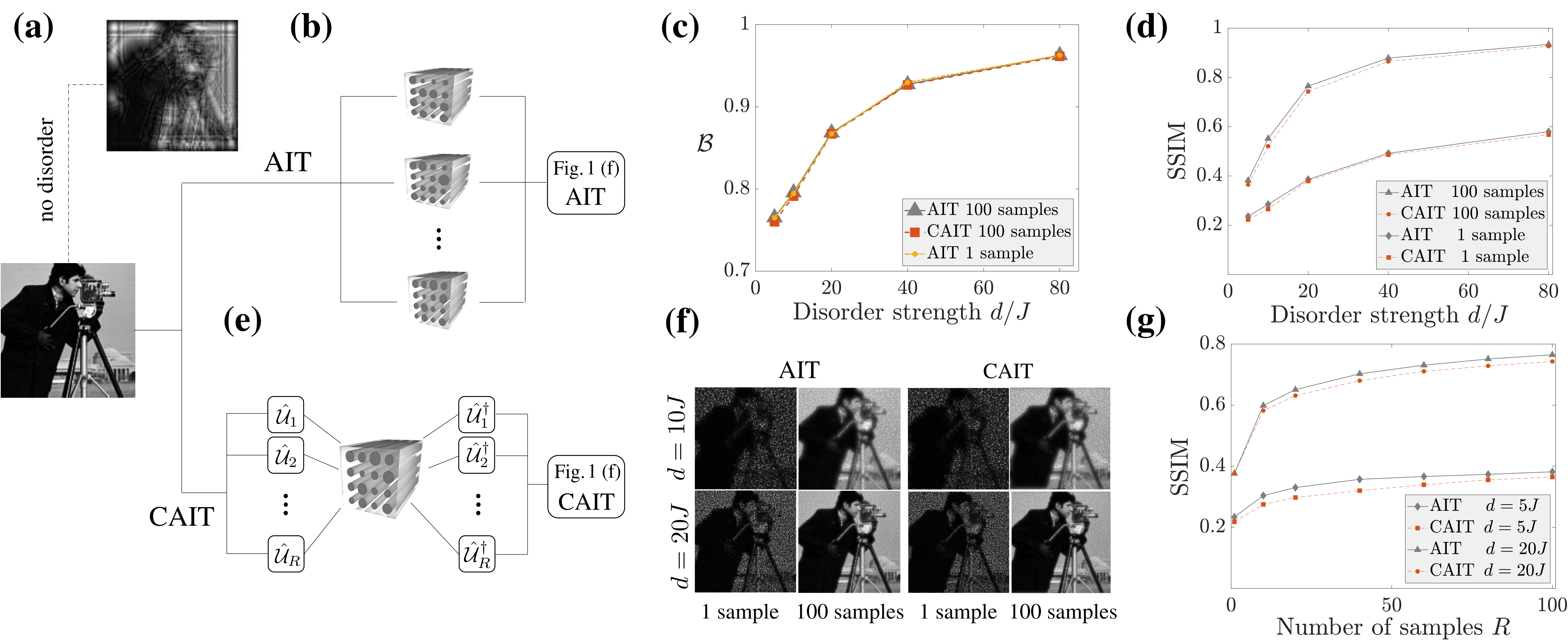}
\vspace{-15pt}
\caption{\ninept \textbf{(a)} Evolved image with no disorder $d=0$. \textbf{(b)} Schematic of the AIT protocol: Sent $R$ copies of $\ket{\psi_{I}}$ through $R$ waveguide lattices with different disorder. The state is then evolved along the transverse direction $\hat{z}$. The magnitudes of the final states $|\bra{j}\ket{\psi_f}|$ for each waveguide are averaged over disorder realizations and we retrieve the final image. (\textbf{(c)} Magnitude overlap $\mathcal{B}$ given by Eq.~(\ref{eq:magniover}) as a function of disorder strength. For both AIT and CAIT protocols, we plot the average over $100$ samples and furthermore we include $1$ sample for the AIT case. \textbf{(d)} SSIM in dependence of disorder strength. Both AIT and CAIT results are plotted for 1 and 100 samples. \textbf{(e)} Schematic of CAIT protocol: Prepare $R$ copies of the initial image and perform a random cyclic permutation on the pixels of each copy as described in the text. Only one disordered waveguide lattice is necessary.  The state is then evolved along the transverse direction and the magnitudes are extracted. Revert the permutation for each time-evolved image and average them in order to acquire the final image. \textbf{(f)} Results of AIT and CAIT over different disorder strengths, and disorder samples. \textbf{(g)} SSIM in dependence of number of samples $R$ for two different disorder strengths. Note that the evolution time for all final images and plots is $t=10/J$.\vspace{-11pt}}
\label{fig:im1}
\end{figure*}
}
 
\subsection{Anderson model}\label{sec:model}
The 2D Anderson Hamiltonian for a square lattice of size $L$, \cite{Giordani2021, sheng2006introduction}, see Fig.\ref{fig:lattice} (a), is given by
\begin{equation}\label{eq:And2d}
    \mathbf{H}=\sum_{l=1}^{L^2} \epsilon_{l}\ket{l}\bra{l}+J\sum^{L^2}_{<l,l'>}\ket{l}\bra{l'}.
\end{equation}
Given a $n\times n$ image, see Sec.~\ref{Imageencoding}, we choose the lattice size $L=n$, where $l=i+n(j-1)$  maps the reshaped state Eq.~(\ref{Init State}), aka the normalized pixel value, to the $i$-th site on a square lattice, see Fig.~\ref{fig:lattice} (a) for a $4\times 4$ example. Furthermore, $\epsilon_{j}$ correspond to the onsite energies (disorder potentials) and $J$ is the hopping amplitude that is restricted to be non-zero only between nearest neighbors, see Fig.\ref{fig:lattice} (a). $\epsilon_{j}$ is drawn from a uniform distribution in the interval $\epsilon_{j} \in \left[-d/2,d/2\right]$, where $d \in \mathbb{R}^{+}$ will be hereafter referred to as the disorder strength. The Hamiltonian can be written in matrix form and satisfies the eigenvalue equation
$\mathbf{H} \ket{\phi_{k}}=E_{k} \ket{\phi_{k}}$, where $E_{k}$ and $\ket{\phi_{k}}$ are respectively the eigenvalues and eigenvectors of the Hamiltonian matrix. According to Anderson's theory, for $N \to \infty$ even the smallest value of disorder strength $d$ will lead to the localization of the eigenvectors in 1D and 2D lattices. Namely, the magnitude profile of the $k$-th eigenvector will decay exponentially
$
\abs{\bra{j}\ket{\phi_{k}}} \sim \mathrm{exp}({-\frac{\abs{j-j_{k}}}{2\xi_{k}}}),
$ 
here $\xi_{k}$ is called the localization length and $j_{k}$ corresponds to the site where the magnitude is peaked, see Fig.~\ref{fig:lattice}(b-c). An important aspect for our discussion is the dependence of the localization length $\xi_{k}$ on the disorder strength. Namely, for two dimensions the localization length decreases exponentially as the strength of disorder is increased \cite{sheng2006introduction}. The regime where our scheme is most efficient, is when the localization length becomes $\xi_{k} \approx 1$. To obtain the disorder strength after which this holds, we numerically extract $\xi_{k}(d)$ for the zero-energy ($E_{k}=0$) eigenvectors, which are known to have the largest localization length \cite{sheng2006introduction}. Our results indicate that we can safely assume we are at the desired regime as long as $d\geq 20J$. The exponential localization of the eigenvectors leads to the suppression of transport along the lattice. This has been experimentally demonstrated in optical waveguide setups \cite{Schwartz2007,Segev2013}.
\subsection{Schrödinger time-evolution}\label{sec:evolution}
The Schrödinger equation is given by
$\mathrm{i}\hbar\frac{\partial}{\partial t} \ket{\psi (t)}= \mathbf{H}\ket{\psi(t)}.$
Furthermore, the time-evolved state at any time $t$ is given by  
$\ket{\psi(t)}= \mathrm{exp}({-\frac{\mathrm{i}}{\hbar}\mathbf{H}t}) \ket{\psi (t=0)},$
where one has to compute the matrix exponential \cite{HOGBEN2011179,KUPROV201131}, which at the same time is the most numerically demanding step in the specified protocol Sec. \ref{sec:results}.

\subsection{Image transfer protocols}\label{sec:results} 
We will now discuss two different protocols for augmenting an image via an Anderson Hamiltonian, which is equivalent to transferring an image over a network of transversely disordered optical waveguides, see Fig.~\ref{fig:im1}(b). Disorder samples $R$ can be taken individually into account or as an average of samples. This, of course, depends on the considered application. The Anderson Image Transfer (AIT) protocol follows a schematic, see Fig.~\ref{fig:im1} (b), where we average the magnitudes of the final state $\left|\bra{j}\ket{\Psi_f}\right|$ over $R$ disorder samples.
\begin{enumerate}
    \item Encode image into quantum state $\ket{\Psi_I}$ Sec.~\ref{Init State}.
    \vspace{-3pt}
    \item Initialize random Hamiltonian $\mathbf{H}$ Sec.~\ref{sec:model}. 
    \vspace{-3pt}
    \item Time-evolve state to time $t$ via Schrödinger equation Sec.~\ref{sec:evolution}.
    \vspace{-13pt}
    \item Save magnitudes of the final state $\ket{\Psi_f}$, reshape back to image Sec.~\ref{Init State}.
\end{enumerate}
 
The second protocol will be referred to as Cyclic Anderson Image Transfer (CAIT). In this case, we initially prepare $R$ copies of the initial image. Each of them is subjected to a different unitary transformation $\hat{\mathcal{U}}_{R}$ that rearranges the pixels of the image by a random cyclic permutation, i.e. the indices $\left[1,2,3,4\right]\rightarrow \left[3,4,1,2\right]$. We again may  consider individual samples or an average over $R$ different cyclic permuted samples. However, each of the rearranged images is then evolved via the same disordered Hamiltonian. By keeping track of the unitary transformations we can then apply the reverse transformation $\hat{\mathcal{U}}_{R}^{\dagger}$ to the evolved states and then average over them in order to acquire the final image. The protocol for CAIT (for schematic, see Fig.~\ref{fig:im1} (e)) follows:
\vspace{-2pt}
\begin{enumerate}
    \item Encode image into quantum state $\ket{\Psi_I}$ Sec.~\ref{Init State}.
    \vspace{-6pt}
    \item Make random cyclic permutation by applying $\hat{\mathcal{U}}_{R}$.
    \vspace{-4pt}
    \item Initialize random Hamiltonian $\mathbf{H}$ Sec.~\ref{sec:model}.
    \vspace{-4pt}
    \item Time-evolve state to time $t$ via Schrödinger equation Sec.~\ref{sec:evolution}.
    \vspace{-14pt}
    \item Revert random cyclic permutation by applying $\hat{\mathcal{U}}_{R}^\dagger$.
    \vspace{-4pt}
    \item Save magnitudes of the final state $\ket{\Psi_f}$, reshape back to image Sec.~\ref{Init State}.
\end{enumerate}
\vspace{-12pt}

\subsection{Similarity measures}\label{sec:similar}
We introduce two quantities to measure the similarity of our augmentation procedure and to understand the effects on a state from a classical and quantum mechanical perspective. 
From the classical perspective, significant efforts \cite{sara2019image} have been made to develop suitable measures, that take into account local structural characteristics of the image, in order to ensure its appearance is maintained during processing. One of the most well established measures for this purpose is the Structural Similarity Index Metrics (SSIM) \cite{wang2004image}. For two image windows $x$,$y$ of the same size, SSIM is defined as:
\begin{equation} \label{eq:ssim}
\mathrm{SSIM}(x,y)=(l(x,y))^{\alpha}(c(x,y))^\beta(s(x,y))^\gamma,
\end{equation}
where $l$ corresponds to luminance, $c$ to contrast, $s$ to correlation and the exponents are usually taken to be $\alpha=\beta=\gamma=1$. The mathematical properties of the SSIM have been examined in detail in \cite{brunet2011mathematical}.
For the quantum perspective we should be aware of the non-locality of quantum systems, thus we consider a global similarity measure, the Bhattacharyya coefficient\cite{fukunaga2013introduction}
\begin{align}\label{eq:magniover}
    \mathcal{B}=\sum_j^{L^2} \sqrt{P_I(j)P_f(j)}=\sum_j^{L^2} \left| \bra{j}\ket{\psi_I}\right|\left| \bra{j}\ket{\psi_f}\right|,
\end{align}
where $P_I(j)= \left| \bra{j}\ket{\psi_I}\right|^2 $ and $P_f(j)= \left| \bra{j}\ket{\psi_f}\right|^2 $. The Bhattacharyya coefficient quantifies the similarity of two probability distributions or here the overlap of initial and final magnitude at sites $j$. This quantity has previously been considered for disordered systems simulated on quantum hardware \cite{Yao2022}. However, since it considers only magnitudes, entanglement encoded in the phases of a state will be neglected. This makes the Bhattacharyya not the best quanitity for all considerations of quantum properties. The augmentation presented here only considers magnitudes, thus $\mathcal{B}$ is a well fitted similarity measure for our case. In the analysis that follows, we will compare and contrast the SSIM and the Bhattacharyya in order to understand the effects of the augmentation schemes.

\section{RESULTS \& DISCUSSION}
\label{sec:discussion}
For simplicity we set $\hbar=1$. Additionally, we express time $t$ in units of $\frac{1}{J}$ and disorder strength $d$ in units of $J$, which physically corresponds to a frequency. Furthermore, we set $t=10\frac{1}{J}$ as localization sets in before and a longer time-evolution is unnecessary. Finally, we set $J\equiv 1$ in all calculations, which reduces the discussion in this paper to two parameters: the disorder strength $d$ and the amount of samples (realizations) $R$. 
The image we chose to demonstrate the effect of the augmentation is a grayscale image, known as camera man \cite{cameraman}, which we crop to  $150 \times 150$ pixels. Mapping this image to a state we end up with a state vector of $22500$ entries and a Hamiltonian matrix of dimensions $22500 \times 22500$.  
If no disorder is imposed on the system (i.e. the diagonal elements of the Hamiltonian matrix are zero) then also no averaging is necessary. The magnitudes of the final state $\ket{\psi_f}$ are mapped back into an image where the quality is significantly deteriorated (e.g. see Fig. \ref{fig:im1} (a)). 
Still, if $d\rightarrow\infty$ then the hopping will be completely suppressed, and it is equivalent to having a Hamiltonian matrix with only diagonal terms. This will not change the magnitudes of the states since the phase is neglected by the absolute value. 
Taking the limit $d\rightarrow \infty$ is technically unreasonable or non-implementable in waveguides. Suitable disorder ranges and localization lengths \cite{Demuth2022} one would expect roughly between $d\in [5,20]$, where $\xi$ will be of order 1 for $d=10$ and $\xi\leq 1$ for $d\geq 20$.
This accessible disorder range suffices for high quality image transfer, as presented in the following. Simply by looking at the results for $R=1$ sample at the given disorder strengths, see Fig.~\ref{fig:im1} (f), one may observe the quality of the transferred image due to disorder strength alone. The enhanced quality after averaging is nicely quantified by the SSIM Fig.~\ref{fig:im1} (d), which in the reasonable disorder range reaches nearly 80\% in SSIM. It can be seen that details of the camera man image are nicely recovered already by 100 disorder realizations. 
The AIT protocol Fig.~\ref{fig:im1} (b), needs however multiple disorder samples of $\mathbf{H}$, which corresponds to a large network of waveguides; in particular, the network needs to have a size of $R$ samples of $150\times 150$ waveguides. This is a problem practically (since it means fabricating many different waveguide samples), as well as computationally demanding. If possible a disorder average should be considered  for optimal results\cite{Segev2013}. Here, our second protocol, the CAIT, may be of greater relevance. Only one disorder realization is necessary,  see Fig.~\ref{fig:im1} (e), but random cyclic permutations need to be imposed on the image and reverted after the transfer through the fiber. The cyclic permutations could be implemented using digital mirror devices. In this protocol the disorder average is completely replaced by the cyclic permutation, which yields nearly indistinguishable results to AIT disorder sampled augmentation, Fig.~\ref{fig:im1} (f). This is again very nicely quantified by the SSIM, Fig.~\ref{fig:im1} (d,g) by reaching values close to 80\% for $d=20J$ and $R=100$. On the other hand, the results of the magnitude overlap $\mathcal{B}$, see Fig.~\ref{fig:im1} (c) do not get better with increasing number of averaging $R$. Even a single realization at the relevant disorder range is close to $\mathcal{B}=88\%$. Our interpretation is that the averaging will not fundamentally change global properties of the state. \\
Therefore, we conclude the following: Both averaging procedures, in the AIT or CAIT protocols, increase the similarity locally as captured by the SSIM, se Fig.~\ref{fig:im1} (d,g). They do not change global properties $\mathcal{B}$, see Fig.~\ref{fig:im1}(c), which can be understood as non-local correlations (common in quantum physics but not exclusive to it\cite{Garisto2022}, and sometimes referred to as the classical part of entanglement). The average technique smoothens the localized profiles locally since single realizations may have magnitude fluctuations, see Fig.~\ref{fig:lattice} (b-c). On the other hand, given that the magnitude overlap $\mathcal{B}$ is a global property, its evaluation yields a natural way of averaging over all localized eigenmodes of $\mathbf{H}$, which has a smoothing effect as well. It makes no difference if we take a disorder or cyclic average procedure, as the average over all localized eigenmodes seems to be equivalent. Our results implicate that the same transversely disordered waveguide setup can be used and instead one should permute the indexes, meaning that we expect a great improvement by the CAIT protocol in image transfer.\\
Another point we want to highlight concerns the time-evolution given via the matrix exponent of the random Hamiltonian, which provides a unitary and random matrix. The cyclic rearrangement of the pixels is also a unitary operation meaning it does not change the basic properties of the matrix, where the unitary property is essential for a quantum implementation. The time-evolution leads to a multiplicative noise which however stays locally, within the averaged localization length $\xi$ and, counter-intuitively, averages (smoothens) out by taking a simple summed average (mean) of the magnitudes of the final states. This kind of noise has been observed previously in setups of ultrasonics and acoustics in disordered materials \cite{andrews2003ultrasonics}. Since similar multiplicative noise has been found due to random scatterings at impurities we suggest that signatures of Anderson localization had been present. To this end, we argue that our protocol will be beneficial for augmentation of wave properties in classical and quantum machine learning. We propose that the multiplicative noise may increase the robustness against input variations in training data which needs to be tested in a seperate contribution.
\section*{Acknowledgements}
N. E. P. acknowledges support by the EU QuantERA Project PACE-IN (GSRT Grant No. T11EPA4-00015). M. K-E. gratefully acknowledge financial support from the DFG through SFB TR 185, project No.277625399. V. F. R., M. T., P. L. and M. K-E. acknowledges support by the Quantum Initiative Rhineland-Palatinate QUIP. N. E. P. and M. K-E. would like to thank M. Fleischhauer and D. Petrosyan for fruitful discussions.
\bibliographystyle{IEEEbib}

\end{document}